\documentclass{JHEP3}

\usepackage{axodraw, epsf}

\def \be {\begin{equation}}
\def \ee {\end{equation}}
\def \bea {\begin{eqnarray}}
\def \eea {\end{eqnarray}}
\def \nn {\nonumber}

\def \a {\alpha}
\def \b {\beta}

\def \d {\delta}

\def \m {\mu}
\def \n {\nu}
\def \k {\kappa}
\def \lam {\lambda}

\def \s {\sigma}
\def \r {\rho}
\def \o {\omega}

\def \th {\theta}
\def \Th {\Theta}

\def \t {\tau}
\def \dag {\dagger}
\def \p {\partial}

\def\bd{\begin{document}}
\def\ed{\end{document}}
\def\nn{\nonumber}
\def\bea{\begin{eqnarray}}
\def\eea{\end{eqnarray}}
\let\bm=\bibitem
\let\la=\label

\def\N{{\cal N}}
\def\sst{\scriptscriptstyle}
\def\thetabar{\bar\theta}
\def\Tr{{\rm Tr}}
\def\one{\mbox{1 \kern-.59em {\rm l}}}

%

\def\a{\alpha}      \def\da{{\dot\alpha}}
\def\b{\beta}       \def\db{{\dot\beta}}
\def\c{\gamma}  \def\C{\Gamma}  \def\cdt{\dot\gamma}
\def\d{\delta}  \def\D{\Delta}  \def\ddt{\dot\delta}
\def\e{\epsilon}        \def\vare{\varepsilon}
\def\f{\phi}    \def\F{\Phi}    \def\vvf{\f}
\def\h{\eta}
\def\k{\kappa}
\def\l{\lambda} \def\L{\Lambda}
\def\m{\mu} \def\n{\nu}
\def\o{\omega}
\def\P{\Pi}
\def\r{\rho}
\def\s{\sigma}  \def\S{\Sigma}
\def\t{\tau}
\def\th{\theta} \def\Th{\Theta} \def\vth{\vartheta}
\def\X{\Xeta}
\def\z{\zeta}
\def\w{\wedge}
\def\u{\underline}
\def\hs{\hspace}


\def\cA{{\cal A}} \def\cB{{\cal B}} \def\cC{{\cal C}}
\def\cD{{\cal D}} \def\cE{{\cal E}} \def\cF{{\cal F}}
\def\cG{{\cal G}} \def\cH{{\cal H}} \def\cI{{\cal I}}
\def\cJ{{\cal J}} \def\cK{{\cal K}} \def\cL{{\cal L}}
\def\cM{{\cal M}} \def\cN{{\cal N}} \def\cO{{\cal O}}
\def\cP{{\cal P}} \def\cQ{{\cal Q}} \def\cR{{\cal R}}
\def\cS{{\cal S}} \def\cT{{\cal T}} \def\cU{{\cal U}}
\def\cV{{\cal V}} \def\cW{{\cal W}} \def\cX{{\cal X}}
\def\cY{{\cal Y}} \def\cZ{{\cal Z}}


\def\ua{\underline{\alpha}} \def\ubb{\underline{\beta}}
\def\ug{\underline{\gamma}}
\def\ub{\underline{\phantom{\alpha}}\!\!\!\beta}
\def\uc{\underline{\phantom{\alpha}}\!\!\!\gamma}
\def\um{\underline{\mu}} \def\un{\underline{\nu}}
\def\ud{\underline\delta}
\def\ue{\underline\epsilon}
\def\una{\underline a}\def\unA{\underline A}
\def\unb{\underline b}\def\unB{\underline B}
\def\unc{\underline c}\def\unC{\underline C}
\def\und{\underline d}\def\unD{\underline D}
\def\une{\underline e}\def\unE{\underline E}
\def\unf{\underline{\phantom{e}}\!\!\!\! f}\def\unF{\underline F}
\def\unm{\underline m}\def\unM{\underline M}
\def\unn{\underline n}\def\unN{\underline N}
\def\unp{\underline{\phantom{a}}\!\!\! p}\def\unP{\underline P}
\def\unq{\underline{\phantom{a}}\!\!\! q}
\def\unQ{\underline{\phantom{A}}\!\!\!\! Q}
\def\unH{\underline{H}}
\def\ul{\underline}

\def\As {{A \hspace{-6.4pt} \slash}\;}
\def\bs {{b \hspace{-6.4pt} \slash}\;}
\def\Ds {{D \hspace{-6.4pt} \slash}\;}
\def\ds {{\del \hspace{-6.4pt} \slash}\;}
\def\ss {{\s \hspace{-6.4pt} \slash}\;}
\def\ks {{ k \hspace{-6.4pt} \slash}\;}
\def\ps {{p \hspace{-6.4pt} \slash}\;}
\def\pas {{{p_1} \hspace{-6.4pt} \slash}\;}
\def\pbs {{{p_2} \hspace{-6.4pt} \slash}\;}


\def\Fh{\hat{F}}
\def\Vh{\hat{V}}
\def\Xh{\hat{X}}
\def\ah{\hat{a}}
\def\xh{\hat{x}}
\def\yh{\hat{y}}
\def\ph{\hat{p}}
\def\xih{\hat{\xi}}

\def\psit{\tilde{\psi}}
\def\Psit{\tilde{\Psi}}
\def\tht{\tilde{\th}}

\def\At{\tilde{A}}
\def\Qt{\tilde{Q}}
\def\Rt{\tilde{R}}
\def\Nt{\tilde{N}}

\def\at{\tilde{a}}
\def\st{\tilde{s}}
\def\ft{\tilde{f}}
\def\pt{\tilde{p}}
\def\qt{\tilde{q}}
\def\vt{\tilde{v}}
\def\nt{\tilde{n}}


\def\delb{\bar{\partial}}
\def\bz{\bar{z}}
\def\bD{\bar{D}}
\def\bB{\bar{B}}


\def\bk{{\bf k}}
\def\bl{{\bf l}}
\def\bp{{\bf p}}
\def\bq{{\bf q}}
\def\br{{\bf r}}
\def\bx{{\bf x}}
\def\by{{\bf y}}
\def\bR{{\bf R}}
\def\bV{{\bf V}}


\def\d{\delta}\def\D{\Delta}\def\ddt{\dot\delta}

\def\p{\partial} \def\del{\partial}
\def\xx{\times}
\def\uno{\mbox{1 \kern-.59em {\rm l}}}

\def\trp{^{\top}}
\def\inv{^{-1}}
\def\dag{{^{\dagger}}}
\def\pr{\prime}

\def\rar{\rightarrow}
\def\lar{\leftarrow}
\def\lrar{\leftrightarrow}

\def\cw{{\cal W}}
\def\cz{{\cal Z}}
\def\tcm{\tilde{\cal M}}
\def\sgn{{\rm sgn}}
\def\sd {d^{4|4}}
\def\lan{\langle}
\def\ran{\rangle}

\title{Tree-level Split Helicity Amplitudes in Ambitwistor Space}
\author{Bin Chen\\
Department of Physics,\\
and State Key Laboratory of Nuclear Physics and Technology,\\
Peking University,\\
Beijing 100871, P.R. China\\
\email{bchen01@pku.edu.cn}}

\author{Jun-Bao Wu\\School of Physics, Korea Institute for Advanced Study\\207-43 Cheongnyangni 2-dong,
Dongdaemun-gu, \\Seoul
130-722, Korea\\ \email{junbaowu@kias.re.kr}}

\date{\today}

\abstract{We study all tree-level split helicity gluon amplitudes
by using the recently proposed BCFW recursion relation and Hodges
diagrams in ambitwistor space. We pick out the contributing
diagrams and find that all of them can be divided into triangles
in a suitable way.  We give the explicit expressions for all of
these amplitudes. As an example, we reproduce the six gluon split
NMHV amplitudes in momentum space.}
\preprint{KIAS-P09018\\ ArXiv:0905.0522[hep-th]}
\newpage
\bd

\section{Introduction}

There are a lot of hidden elegant structures in tree-level
amplitudes in Yang-Mills theory. These structures cannot be seen
from the standard Feynman rules directly. One typical example is the
Parke-Taylor formula for the MHV amplitudes \cite{ParkeTaylor} which
was later proved in \cite{MHV2}. In 2003, by transforming the
amplitudes into twistor space, Witten found a beautiful explanation
of the simplicity of these amplitudes\cite{Witten:2003nn}. He
furthermore proposed a dual topological B model in super twistor
space ${\bf CP}^{3|4}$ to calculate the tree-level amplitudes. Being
different from the usual strong-weak duality in AdS/CFT
correspondence, his proposal is a kind of weak-weak duality.
Moreover, the study of this twistor string theory inspired the
development of a new computation formalism - MHV diagrams \cite{CSW}
(also known as CSW rules) in which off-shell continuations of MHV
amplitudes is used as vertices to build up the diagrams. Later on an
on-shell recursion relation was proposed in \cite{BCF, BCFW}. This
so-called BCFW recursion relation is a very powerful computation
tool
at tree level. 
Last year, all tree-level amplitudes in ${\cal N}=4$ super
Yang-Mills theory and ${\cal N}=8$ supergravity were computed in
\cite{DrummondHenn} and \cite{DSVW} by using the maximal
supersymmetric version \cite{Brandhuber:2008pf,
ArkaniHamed:2008gz} of BCFW relation.

Recently, it was revealed that there were more surprises in the
amplitudes of (super-)Yang-Mills and (super-)gravity in
(ambi)-twistor space\cite{Mason:2009sa,ACCK}. Especially, in
\cite{ACCK} it was shown that all tree amplitudes could be combined
into an ``S-matrix" scattering functional in twistor space. The
on-shell (super-)BCFW relation in ambitwistor space (using both
twistor and dual twistor) plays an essential role in their study. In
the (ambi-)twistor space, the relation turns out to be simpler and
more elegant. In \cite{ACCK}, a diagrammatic representation was also
given for this recursion relation. Using the BCFW formula, the
multiparticle amplitudes could be represented by a set of
diagrammatic rule, which actually gave a concrete realization of
Penrose's ``twistor diagram program" \cite{Penrose:1972ia,
Hodges:1980hn}. In fact, the twistor diagram formalism  in
\cite{ACCK} is a refined version of the formalism developed by
Hodges in \cite{Hodges:2005bf, Hodges:2005aj, Hodges:2006tw}. It
turns out that considering amplitudes in (ambi-)twistor space could
not only make the computation
 much easier but also uncover the underlying structure in Yang-Mills theory.

Among all of the tree-level gluon (partial) amplitudes, the class
of split helicity amplitudes is of particular interest. In these
amplitudes, the gluons with the same helicity are put together.
The amplitudes in this class are closed under the original BCFW
recursion relations for gluon amplitudes \cite{BCF}. These
amplitudes have been computed in momentum space \cite{BFRSV} by
solving these recursion relations. In \cite{Drummond:2008vq}, it
has been shown that they are also dual conformal covariant. In
this paper, we will study these amplitudes in ambitwistor space.

The main tools we will use are the BCFW relation and the Hodges
diagrams developed in \cite{ACCK}. One of the key points in our
computation is the following: due to the Grassman integration used
to pick out the gluon amplitude from the superamplitudes, many
Hodges diagrams do not really contribute so that they could be
thrown away. This idea simplifies the computation significantly. By
investigating some simple examples, we come to the conclusion that
each of the remaining Hodges diagram can be divided into triangles
in a suitable way. These triangles are naturally combined into some
domains. Besides the MHV ($\overline{MHV}$) domains discussed in
\cite{talkNima, talkKaplan}, there is another new kind of domains
which we name as ``type II'' domains. One of the differences between
these two kinds of domains is the helicities of the external gluon
of the domains. Using these domains, we can write down all of the
expressions for the split helicity amplitudes. We also check that
our result reproduce the six gluon split helicity next-to-MHV (NMHV)
amplitude.

In the next section of this paper, we will give a brief
introduction to the BCFW recursion relation in ambitwistor space
and the Hodges diagrams in \cite{ACCK}. We will study the split
helicity amplitudes in section~\ref{sec3}. After some general
discussions in subsection~\ref{ss31}, we will first study split
MHV amplitudes in subsection~\ref{ss32}. We then study two
examples with few external gluons in subsection~\ref{ss33}. We
give our general results for split helicity amplitudes in
subsection~\ref{ss34}. In section~\ref{sec4}, we show that we
reproduce the six-gluon split helicity NMHV amplitudes in momentum
space. The final section is devoted to conclusion and discussions.

\section{BCFW recursion relation in ambitwistor space}

The on-shell momentum of external gluon $p^\mu$ can be expressed
in terms of two spinors $\lambda_\alpha,
\tilde\lambda_{\dot{\alpha}}$ as follows: \be
p_\mu\sigma^\mu_{\a\dot{a}}=\lam_{\a}\tilde\lam_{\dot{a}}. \ee
Since we compute the amplitudes in the space with $(2, 2)$
signature, $\lam$ and $\tilde\lam$ are independent real spinors.
For a function $f$ of $\lambda, \tilde\lam$, we can freely
transform it into twistor space: \be f(W)=\int d^2\lam
\exp[i\tilde\mu^\a\lambda_\a]f(\lam, \tilde\lam),
\hs{5ex}W_A=(\tilde\mu, \tilde\lam) \ee or into dual twistor
space: \be f(Z)=\int d^2\tilde\lam
\exp[i\mu^{\dot{\a}}\tilde\lam_{\dot{\a}}]f(\lam, \tilde\lam),
\hs{5ex}Z^A=(\lam, \mu). \ee The following combinations appear
almost everywhere in the amplitudes in ambitwistor space: \be
W\cdot Z\equiv\tilde\mu\lambda-\mu\tilde\lambda,\hs{3ex}
W_iIW_j\equiv[\tilde\lambda_i, \tilde\lambda_j],\hs{3ex}
Z_iIZ_j\equiv\langle\lam_i, \lam_j\rangle, \ee where $I^{AB}$ and
$I_{AB}$ are ``infinity twistors".

In ${\cal N}=4$ super Yang-Mills theory, all states in the vector
multiplet can be obtained by the supersymmetry transformation
acting on the gluon with helicity $-1$ (or $+1$). Using this fact
we can introduce a on-shell superspace \cite{Nair:1988bq,
Georgiou:2004by, Bianchi:2008pu, Brandhuber:2008pf,
ArkaniHamed:2008gz} with Grassman coordinates $\eta_I$ (or
$\tilde\eta_I$), $I=1, \cdots, 4$. The gluon with helicity $-1$
(or $+1$) is related to the state in the superspace with $\eta=0$
($\tilde\eta=0$). Thus we can lift the amplitudes in momentum
space to a superamplitudes, $M(\lambda, \tilde\lambda, \eta)$ (or
$M(\lambda, \tilde\lambda, \tilde\eta)$). After performing the
expansion of $\eta_i$, the term with no $\eta_i$'s gives the
amplitudes with the $i$-th particle being gluon with helicity
$-1$, while the coefficient of $\eta_i^4$  gives the amplitudes
with the $i$-th particle being gluons with helicity $+1$. The
amplitudes with the particles being gluinos or scalars come from
the terms with other products of $\eta_i$'s. We have a similar
result for the expansion of $\tilde\eta_i$. Notice that for each
particle, we can choose either $\eta$ or $\tilde\eta$ but not
both.

We choose $\tilde\eta$ in the twistor space and combine it with $W$
into a supertwistor: \be \cw=(W_A, \tilde\eta^I).\ee Similarly for
the dual twistor, we have \be \cz=(Z^A, \eta_I).\ee Now the
superamplitudes are the functions of $\cw_i$'s and $\cz_j$'s. And
similarly in the superamplitudes we always have the following
combinations: \bea \cw\cdot\cz=W\cdot Z+\eta\cdot\tilde\eta,\hs{3ex}
\cw_iI\cw_j=W_iIW_j,\hs{3ex} \cz_iI\cz_j=W_iIZ_j.
\label{supertwistors}\eea If we choose $\cw_i$ ($\cz_i$) for the
$i$-th external particle, the amplitudes $M$ be with weight $2h_i-2$
($-2h_i-2$) under the scaling transformation $\cw_i\to t\cw_i$
($\cz_i\to t\cz_i$) \cite{ACCK}.

\begin{figure}[ht!]
    \epsfxsize=100mm%
    \hfill\epsfbox{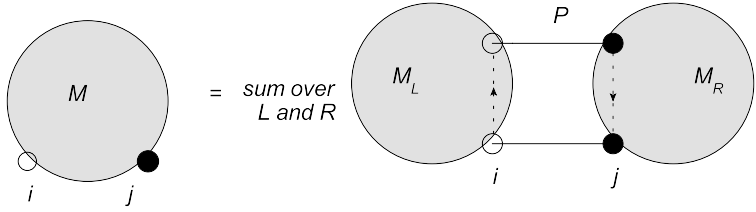}\hfill~\\
    \caption{Hodges diagram for the BCFW relations.}
  \label{fig1}
   \end{figure}

    \begin{figure}[ht!]
    \epsfxsize=100mm%
    \hfill\epsfbox{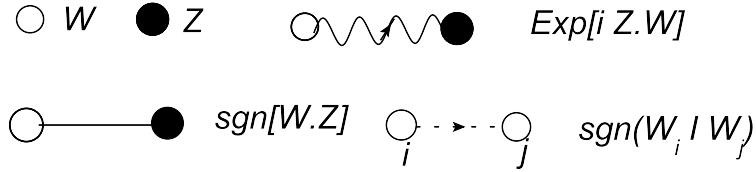}\hfill~\\
    \caption{Some ingredients in Hodges diagrams.}
    \label{fig2}
   \end{figure}

In this paper we will compute the amplitudes using the BCFW relation
in super-ambitwistor space \cite{ACCK}: \bea M(\cw_i,
\cz_j)=\sum_{L, R}\int \large[ D^{3|4}\cw_P D^{3|4}\cz_P
\large]_{\cw_i, \cz_j}M_L(\cw_i; \cw_P)M_R(\cz_j; \cz_P).
\label{bcfw1} \eea Here the projective measure is defined as: \bea
\large[ D^{3|4}\cw_P D^{3|4}\cz_P \large]_{\cw_i,
\cz_j}&=&D^{3|4}\cw_P
D^{3|4}\cz_P\,\sgn(\cw_i\cdot\cz_j)\sgn(\cw_P\cdot\cz_P)\nn\\&\times&\sgn(\cw_i
I \cw_P)\sgn(\cz_P I\cz_j) \eea The projective measure can be
de-projectivized as: \bea M(\cw_i, \cz_j)&=&\sum_{L, R}\int
d^{4|4}\cw d^{4|4}\cz
{\rm sgn}(\cw_i\cdot\cz_j) {\rm sgn}(\cw\cdot\cz)\exp[i\cw_i I \cw] \nn\\
& & \exp[i \cz I \cz_j] M_L(\cw_i, \cw)M_R(\cz_j, \cz). \label{acck}
\eea In \cite{ACCK}, the above relation is presented using Hodges
diagram as in Fig.~\ref{fig1}. Some ingredients in Hodges diagrams
are listed in Fig.~\ref{fig2}. Notice that we add an arrow to the
dash line comparing with \cite{ACCK} since $\sgn(\cw_i I \cw_j)$ is
anti-symmetric. We also add an arrow to the wavy line. The reason is
that, as discussed in \cite{ACCK}, the factor $\exp[i\cz\cdot\cw]$
usually appears with integration over $\cz$ or $\cw$ to represent a
super-twistor transformation (a super-Fourier transformation). We
use the arrow toward $\cz$ (or $\cw$) to represent integration over
$\cz$ (or $\cw$). We find that this is quite useful to make things
clear in the complicated diagrams we will meet later.

\section{Computations of split helicity amplitudes in ambitwistor space\label{sec3}}
\subsection{General discussions on split helicity amplitudes\label{ss31}}

Now we begin our study of the split helicity amplitude $M(1, +;
\cdots, q, +; (q+1), -; \cdots, n, -)$ in ambitwistor space.
We first make the following choice between $W$ and $Z$ for the
external gluons: \be M(W_1, +;\cdots, W_q, +; Z_{q+1}, -; \cdots,
Z_n, -),\ee since with this choice the amplitudes have weight zero
with respect to independent rescalings of each $W$ or $Z$. Notice
that at tree level, the pure gluonic amplitudes in pure Yang-Mills
theory are the same as the ones in ${\cal N}=4$ super Yang-Mills
theory. Now we lift these amplitudes to the super-amplitudes
$M(\cw_1, \cdots, \cw_q, \cz_{q+1}, \cdots, \cz_n)$. The split
helicity gluonic amplitudes can be obtained from the
super-amplitudes through the following Grassman integral: \bea &&
M(W_1, +;\cdots, W_q, +; Z_{q+1}, -; \cdots, Z_n, -)=\int
d^4\tilde\eta_1\cdots d^4\tilde\eta_q
d^4\eta_{q+1}\cdots d^4\eta_n   \nn\\
&&\qquad \qquad \qquad\qquad\qquad \prod_{j=1}^q(\tilde\eta_j)^4
\prod_{j=q+1}^n(\eta_j)^4 M(\cw_1, \cdots, \cw_q, \cz_{q+1},
\cdots, \cz_n). \label{superamp}\eea

To compute the split helicity amplitudes, we do not need to
completely compute the superamplitudes. Instead we pick out the
Hodges diagrams contributing to these special gluonic amplitudes,
by doing Grassman integrals. This play a very crucial role in our
computations.

For the superamplitudes $M(\cw_1, \cdots, \cw_q, \cz_{q+1}, \cdots,
\cz_n)$, choosing $i$ and $j$ in eq.~(\ref{bcfw1}) to be $1$ and $n$ respectively,
the BCFW relation now becomes:
 \bea & &
M(\cw_1, \cdots, \cw_q, \cz_{q+1}, \cdots, \cz_n)=
 \int \large[ D^{3|4}\cw_P D^{3|4}\cz_P \large]_{\cw_1, \cz_n}\nn\\
& &\qquad \qquad (\sum_{i=2}^q M_L(\cw_1, \cdots, \cw_i, \cw_P)M_R(\cz_P, \cw_{i+1}, \cdots, \cw_q, \cz_{q+1}, \cdots, \cz_n)\nn\\
&& \qquad \qquad +\sum_{i=q+1}^{n-2}  M_L(\cw_1, \cdots, \cw_q,
\cz_{q+1} \cdots, \cz_i, \cw_P)M_R(\cz_P, \cz_{i+1}, \cdots,
\cz_n) ). \label{bcfw2} \eea
 Using eq.~(\ref{superamp}) and eq.~(\ref{bcfw2}), we get:
\bea && M(W_1, +;\cdots, W_q, +; Z_{q+1}, -; \cdots, Z_n, -)= \int
d^4\tilde\eta_1
d^4\eta_n(\tilde\eta_1)^4(\eta_n)^4\nn\\
&&\qquad \qquad \int\large[ D^{3|4}\cw_P D^{3|4}\cz_P
\large]_{\cw_1, \cz_n} \left(\sum_{i=2}^q \int \prod_{j=2}^i
d^4\tilde\eta_j (\tilde\eta_j)^4 M_L(\cw_1, \cdots,
\cw_i, \cw_P)\right.\nn\\
&& \qquad \qquad\int \prod_{j=i+1}^q d^4\tilde\eta_j (\tilde
\eta_j)^4 \prod_{j=q+1}^{n-1} d^4\eta_j (\eta_j)^4 M_R(\cz_P,
\cw_{i+1},
\cdots, \cw_q, \cz_{q+1}, \cdots, \cz_n)  \nn\\
&& \qquad \qquad+\sum_{i=q+1}^{n-2}\int \prod_{j=2}^q
d^4\tilde\eta_j (\tilde\eta_j)^4 \prod_{j=q+1}^i d^4\eta_j
\eta_j^4 M_L(\cw_1,
\cdots, \cw_q, \cz_{q+1}, \cdots, \cz_i, \cw_P)\nn\\
& &\qquad \qquad\left. \int\prod_{j=i+1}^{n-1}d^4\eta_j\eta_j^4
M_R(\cz_P, \cz_{i+1}, \cdots, \cz_n)\right). \eea Since we want to
use these Grassman integrals to pick out the needed Hodges
diagrams, we also need to deal with the integration over
$\tilde\eta_1$ and $\eta_n$ \footnote{This needs to be done
separately since the measure in the projective twistor spaces
depends on $\cw_1, \cz_n$.}. To do this, let us consider the more
general integral: \bea & &\int d^4\tilde\eta_1 d^4\eta_n
\tilde\eta_1^4 \eta_n^4 \int \large[ D^{3|4}\cw_P D^{3|4}\cz_P
\large]_{\cw_1, \cz_n} F(\cw_P, \cz_P), \eea where $F(\cw, \cz)$
is a function of weight zero.  As before, we can de-projectivize
the projective measure:
\bea & &\int d^4\tilde\eta_1 d^4\eta_n \tilde\eta_1^4 \eta_n^4 \int \large[ D^{3|4}\cw_P D^{3|4}\cz_P \large]_{\cw_1, \cz_n} F(\cw_P, \cz_P)\nn\\
&&\qquad \qquad =\int d^4\tilde\eta_1 d^4\eta_n \tilde\eta_1^4
\eta_n^4\int
D^{4|4}\cw D^{4|4}\cz {\rm {\rm sgn}}(\cw_1\cdot \cz_n)\nn\\
&& \qquad \qquad \qquad {\rm sgn}(\cw\cdot\cz) \exp(i\cw_1 I
\cw)\exp(i \cz I\cz_n) F(\cw, \cz). \eea By using  \be {\rm
sgn}(x)=\int \frac{da}a \exp(iax),\label{sgn}\ee we get
 \bea & & \int d^4\tilde\eta_1 d^4\eta_n \tilde\eta_1^4
\eta_n^4 \int \large[ D^{3|4}\cw_P D^{3|4}\cz_P \large]_{\cw_1,
\cz_n} F(\cw_P, \cz_P)\nn \\
&&\qquad \qquad =\int D^{4|4}\cw D^{4|4}\cz ({\rm sgn}(\cw_1\cdot
\cz_n) {\rm sgn}(\cw\cdot\cz) \exp(i\cw_1 I \cw)\exp(i \cz
I\cz_n)\nn\\ & & \qquad \qquad \qquad F(\cw, \cz))|_{\cw_1\to W_1,
\cz_n\to Z_n}.\eea So in eq.~(\ref{bcfw2}), performing the
integration: \be  \int d^4\tilde\eta_1 d^4\eta_n \tilde\eta_1^4
\eta_n^4\ee is just performing the replacement: \be \cw_1\to W_1,
\cz_n\to Z_n \ee as usual, and the projective measure will not
affect this.

Now consider $M_L$ in the first sum in eq.~(\ref{bcfw2}), it
appears in the following way: \bea  \int \prod_{j=2}^i
d^4\tilde\eta_j \tilde\eta_j^4 M(\cw_1, \cdots, \cw_i,
\cw_P)|_{\cw_1\to W_1}=M(W_1, +, \cdots, W_i, +, \cw_P)
 \eea

By writing this superamplitudes as linear combination of amplitudes,
we can see that the above result vanishes when $i>2$. So only the
term with $i=2$ contributes to the first sum of eq.~(\ref{bcfw2}).
Similarly only term with $i=n-2$  in the second sum contributes. As
a result for $n>4$, only two terms have nonvanishing contribution,
while for $n=4$ only one term contributes.
Use this we can throw away many Hodges diagrams which will not
contribute to the split helicity amplitudes. This simplifies the
computations significantly.

Two other useful relations are: \bea  \int d^4\tilde\eta_1
d^4\tilde\eta_2\tilde\eta_1^4\tilde\eta^4_2 M(\cw_1, \cw_2, \cw)
&=&M^{++-}(W_1, W_2, W)\tilde\eta^4\nn\\
&=&\int d^4\tilde\eta_1
d^4\tilde\eta_2\tilde\eta^4_1\tilde\eta^4_2 M^+(\cw_1, \cw_2,
\cw),
 \label{symp}\\
  \int
d^4\eta_{n-1} d^4\eta_n \eta^4_{n-1}\eta^4_n M(\cz_{n-1}, \cz_n,
\cz) &=& M^{--+}(Z_1, Z_2, Z) \eta^4\nn\\&=&\int d^4\eta_{n-1}
d^4\eta_n \eta^4_{n-1}\eta^4_n M^-(\cz_{n-1}, \cz_n, \cz),
\label{symn} \eea
where \be M^+(\cw_1, \cw_2, \cw)=\int\sd\cz
e^{i\cw\cdot\cz}\sgn(\cw_1I\cw_2)
\sgn(\cw_1\cdot\cz)\sgn(\cw_2\cdot\cz), \label{3ptp}\ee \be
M^-(\cz_{n-1}, \cz_n, \cz)=\int\sd\cw
e^{i\cz\cdot\cw}\sgn(\cz_{n-1}I\cz_n)
\sgn(\cz_{n-1}\cdot\cw)\sgn(\cz_n\cdot\cw), \label{3ptn} \ee are
given in \cite{ACCK}. The first lines of eq.~(\ref{symp}) can be
understood as follows: the integration over $\tilde\eta$'s fix the
helicity of $\cw_1$ and $\cw_2$, then the helicity of $\cw$ is
fixed to be $-1$ for the amplitude to be nonzero. The fact that,
in eq.~(\ref{3ptp}) (eq.~(\ref{3ptn})), only $M^+$ ($M^-$)
contributes can also be obtained from the ``vanishing'' identity
in \cite{ACCK}.

\begin{figure}[ht]
    \epsfxsize=130mm%
    \hfill\epsfbox{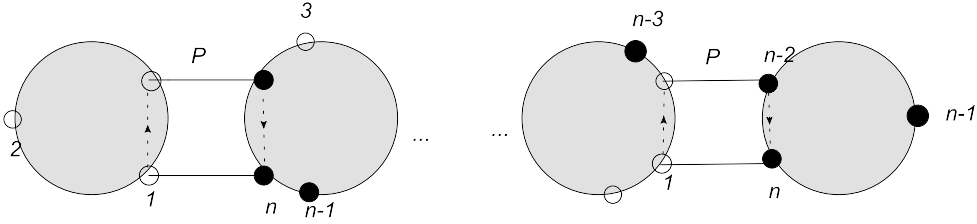}\hfill~\\
    \caption{The contributing Hodges diagrams.}
    \label{fig3}
   \end{figure}

The above discussions lead to the following results for the
amplitudes with $n>4$ gluons:
\bea && M(W_1, +;\cdots, W_q, +; Z_{q+1}, -; \cdots, Z_n, -)= \int
d^4\tilde\eta_1
d^4\eta_n(\tilde\eta_1)^4(\eta_n)^4\nn\\
&&\qquad \qquad \int\large[ D^{3|4}\cw_P D^{3|4}\cz_P
\large]_{\cw_1, \cz_n} \left(\int d^4\tilde\eta_2(\tilde\eta_2)^4
M^+(\cw_1, \cw_2, \cw_P) \right.\nn\\
&&\qquad \qquad \int \prod_{j=i+1}^q d^4\tilde\eta_j (\tilde
\eta_j)^4 \prod_{j=q+1}^{n-1} d^4\eta_j (\eta_j)^4 M_R(\cz_P,
\cw_{i+1}, \cdots, \cw_q, \cz_{q+1}, \cdots, \cz_n)+\nn\\
&&\qquad \qquad \int \prod_{j=2}^q d^4\tilde\eta_j
(\tilde\eta_j)^4 \prod_{j=q+1}^{n-2} d^4\eta_j \eta_j^4 M_L(\cw_1,
\cdots, \cw_q, \cz_{q+1}, \cdots, \cz_{n-2},
\cw_P) \nn\\
&&\qquad \qquad\left.\int d^4\eta_{n-1}\eta_{n-1}^4 M^-(\cz_P,
\cz_{n-1}, \cz_n)\right).\label{3pt}\eea The result can be
expressed as in Fig.~\ref{fig3}.

The BCFW relation in the super-twistor space treat the whole ${\cal
N}=4$ multiplet once at a time by using the on-shell ${\cal N}=4$
superspace. As we can see from the above, fixing the helicities of
the external gluon fixes the helicities of some internal particles
in some Hodges diagrams. This in turn fixes some helicities of other
internal particles. The reason is that as the BCFW relation for
gluonic amplitudes in momentum space, the helicity of $\cw_P$ and
$\cz_P$ should be opposite for each term in the expansion of the
superamplitudes in the right-hand-side of eq.~(\ref{bcfw1}). To see
this, using eqs.~(\ref{supertwistors}, \ref{acck}, \ref{sgn}), we
can pick out the integral over $\eta$ and $\tilde\eta$ in $M$, \be
\int d^4\eta d^4\tilde\eta \exp[i a
\eta\cdot\tilde\eta]M_L(\tilde\eta)M_R(\eta). \ee If we consider the
terms in $M_L$ whose $\cw_P$ has helicity $+1$ ($-1$), then these
terms will have no $\tilde\eta$'s (four $\tilde\eta$'s). From the
above equation, we can see that the terms in $M_R$ with no $\eta$'s
(four $\eta$'s) are picked out, which means that the helicity of
$\cz_P$ have to be $-1$ ($+1$). (This result is also valid if we
exchange ($\cw_P$,  $\tilde\eta$) and ($\cz_P$,
$\eta$).)\footnote{Similar discussions also tell us if one of the
two helicities of two nods linked by the wavy line is fixed, the
other will be automatically fixed to be the same.} We have seen that
in the first term of eq.~(\ref{3pt}), the helicity of $\cw_P$ is
fixed by eq.~(\ref{symp}) to be $-1$. Then the helicity of $\cz_P$
in this term is fixed to be $+1$, the same as the helicity of
$\cw_2$.  This guarantees that we can treat the remaining $M_R$ as a
split helicity amplitude similar to the one we begin with and we can
still throw away many Hodges diagrams when we compute it using BCFW
rules as the first step. We have similar results for the second term
in eq.~(\ref{3pt}).

\begin{figure}[ht]
    \epsfxsize=140mm%
    \hfill\epsfbox{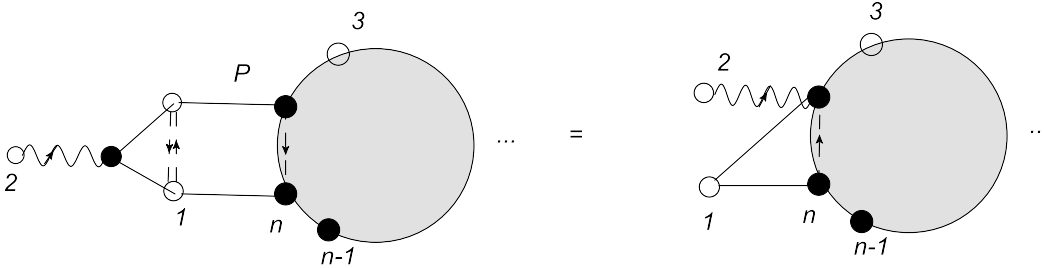}\hfill~\\
    \caption{Simplifying the left graph in Fig.~3 using various identities.}
    \label{fig4}
   \end{figure}

\begin{figure}[ht]
    \epsfxsize=50mm%
    \hfill\epsfbox{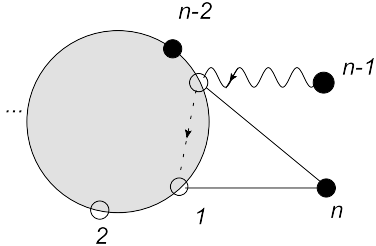}\hfill~\\
    \caption{Simplifying the right graph in Fig.~3 using various identities.}
    \label{fig5}
   \end{figure}

Now we can use the Hodges diagram for three gluon amplitudes
$M^{\pm}_{sym}$, $\sgn^2=1$ and ``scrunch'' identity in
\cite{ACCK} to simplify the graphs in Fig.~\ref{fig3}. The first
graph is simplified as Fig.~\ref{fig4}, while the second graph is
simplified to be Fig.~\ref{fig5}. We can think Fig.~\ref{fig4} and
Fig.~\ref{fig5} as the simplified recursion relations in this
special case.

\subsection{A special case: split MHV amplitudes\label{ss32}}

   \begin{figure}[ht]
    \epsfxsize=120mm%
    \hfill\epsfbox{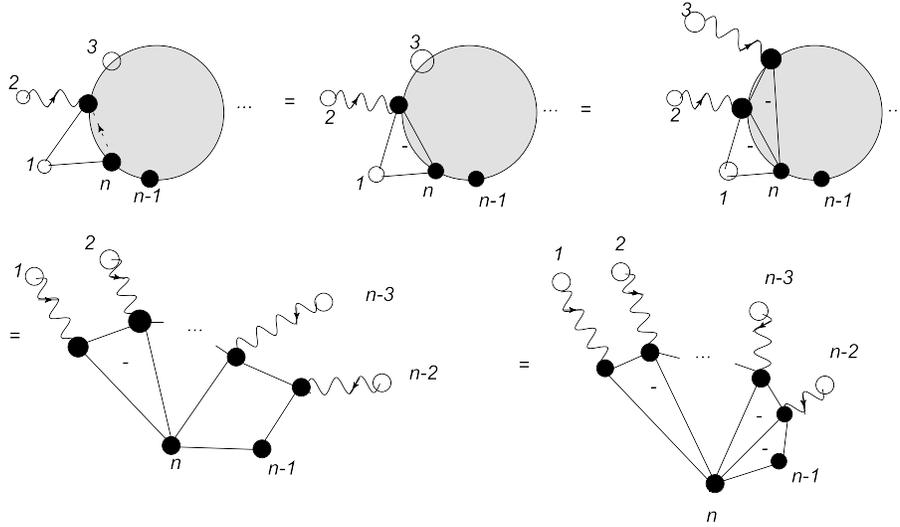}\hfill~\\
    \caption{The split MHV amplitudes.}
  \label{fig6}

   \end{figure}
      \begin{figure}[ht]
    \epsfxsize=100mm%
    \hfill\epsfbox{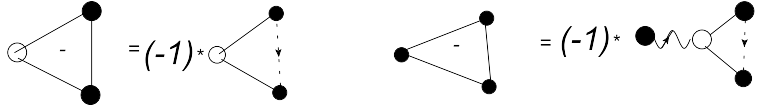}\hfill~\\
    \caption{The triangles.}
    \label{fig6b}
   \end{figure}
Let us now have a close look at the split MHV amplitudes because
they are the simplest and later we will find that they will be one
of the building blocks of the general split helicity amplitudes. Let
us consider the amplitudes \be M(W_1, +;\cdots, W_{n-2}, +; Z_{n-1},
-; Z_n, -).\ee Using the arguments in previous section, we can
perform the calculations as in Fig.~\ref{fig6}. We use the triangles
with signs to represent $-M^{\pm}_3$ (Fig.~\ref{fig6b}), as in
\cite{talkNima, talkKaplan}. Notice that in this special case, at
every step, we only have one diagram since the other one vanishes
due to the vanishing of the subamplitude $M^{+\cdots+-}$. Here we
have also used the results for four-gluon MHV amplitudes in
\cite{ACCK}.

   \begin{figure}[ht]
    \epsfxsize=100mm%
    \hfill\epsfbox{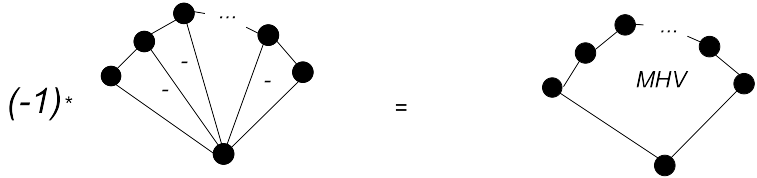}\hfill~\\
    \caption{The MHV domain.}
    \label{fig7}
   \end{figure}

 \begin{figure}[ht]
    \epsfxsize=100mm%
    \hfill\epsfbox{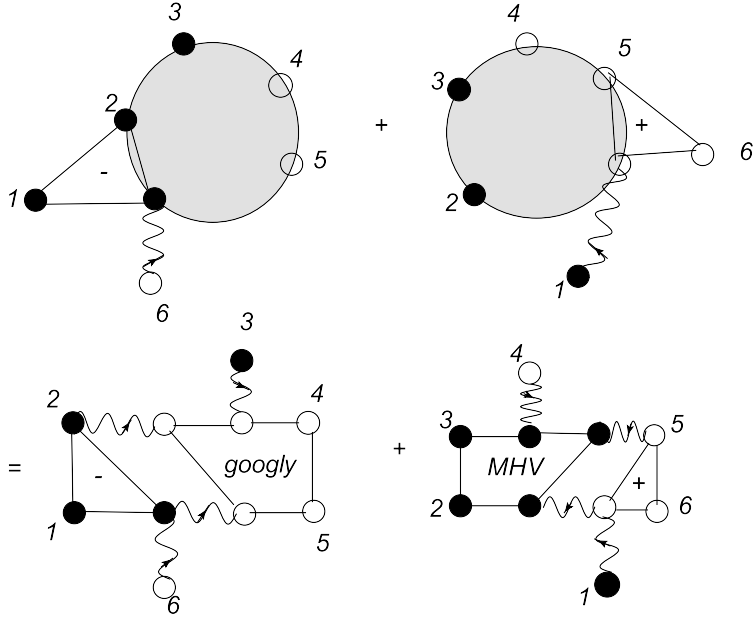}\hfill~\\
    \caption{A NMHV six-gluon amplitude.}
    \label{fig8}
   \end{figure}

   \begin{figure}[ht]
    \epsfxsize=150mm%
    \hfill\epsfbox{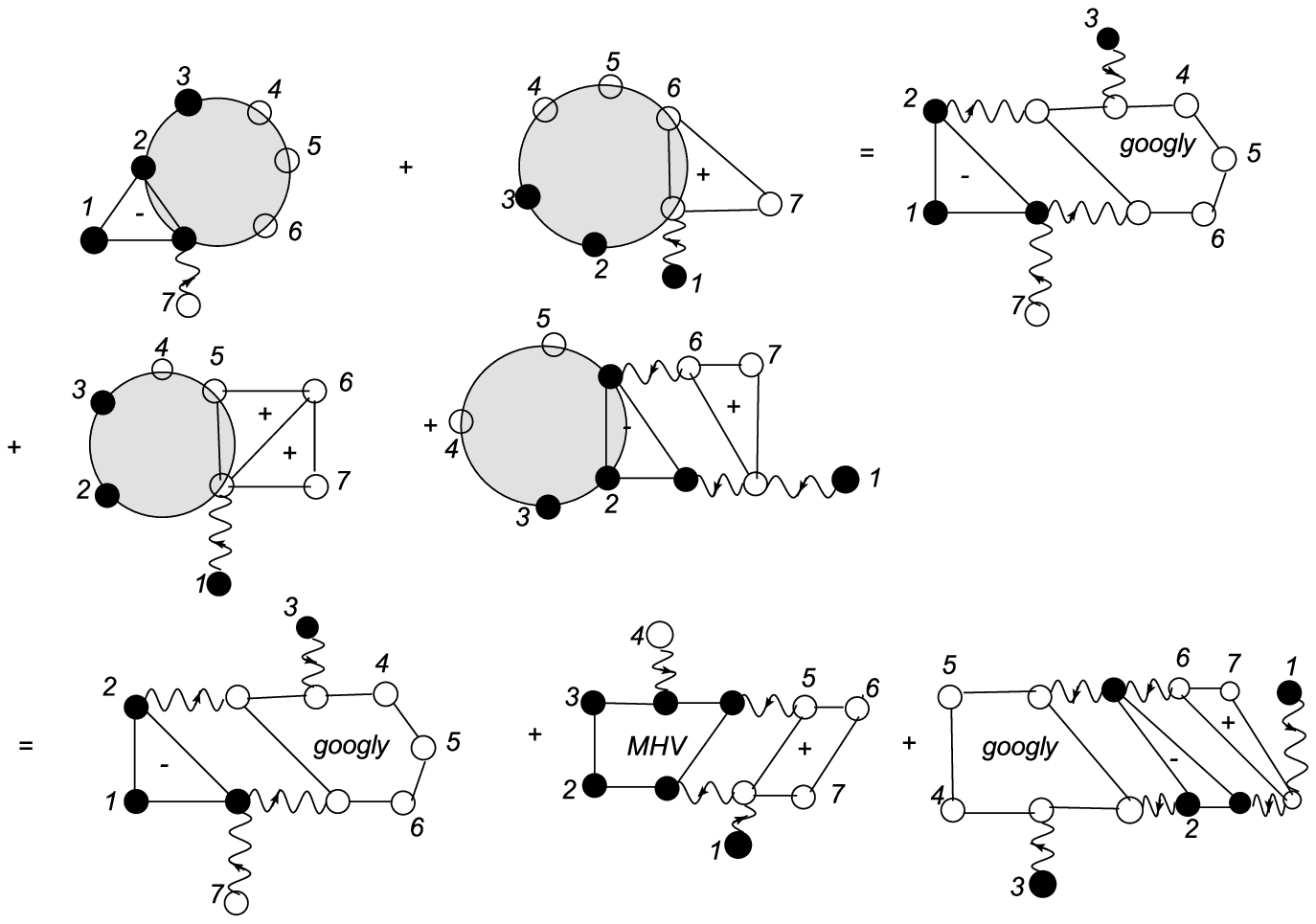}\hfill~\\
    \caption{A NNMHV seven-gluon amplitude.}
    \label{fig9}
   \end{figure}

From this result, we can see that the more natural choice of
$\cw$'s and $\cz$'s is to choose $\cz$'s for all of the external
gluons. We
can draw the diagram as in Fig.~\ref{fig7}. We find that the triangles are combined into a  
MHV-domains \cite{talkNima, talkKaplan}. Similarly we can get the
$\overline{MHV}$(googly)-domain, by changing all $\cz$'s into $\cw$'s and
all black nods into white nods.

\subsection{Two non-MHV examples\label{ss33}}

\begin{figure}[ht]
    \epsfxsize=120mm%
    \hfill\epsfbox{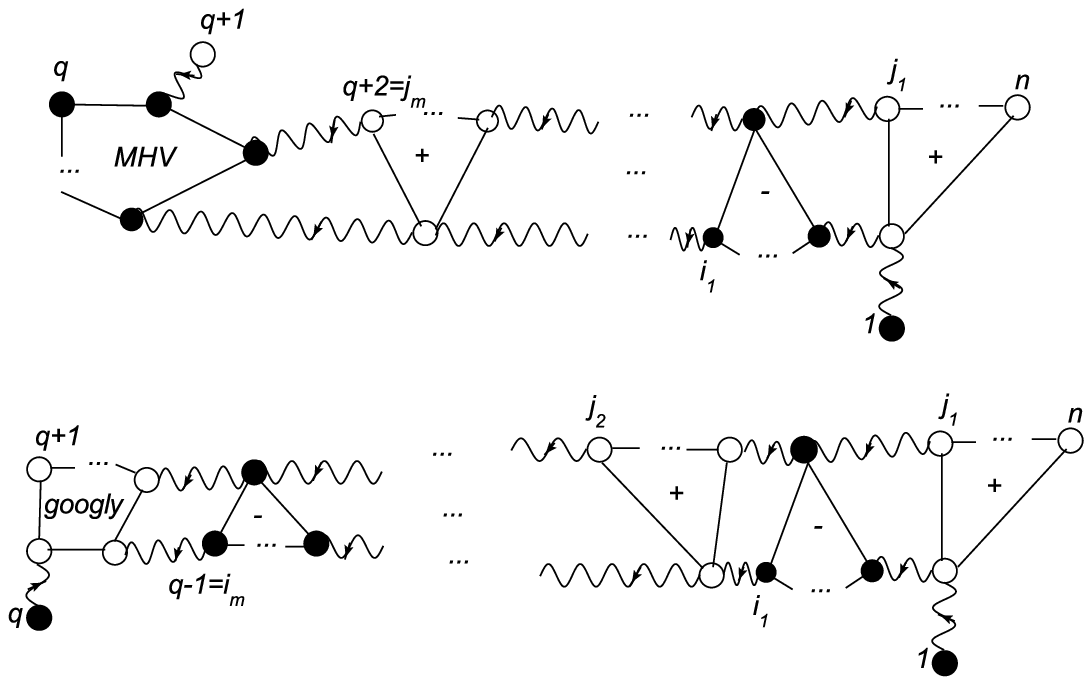}\hfill~\\
    \caption{The diagrams for generic split helicity amplitudes: part I.}
    \label{fig10}
   \end{figure}

   \begin{figure}[ht]
    \epsfxsize=140mm%
    \hfill\epsfbox{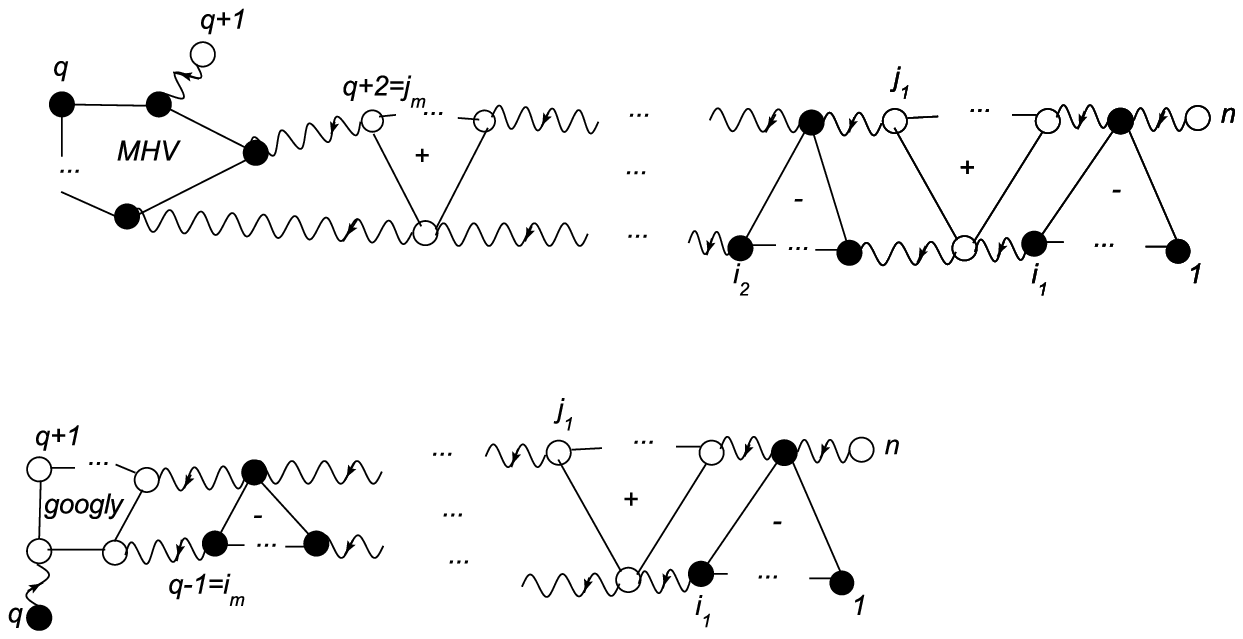}\hfill~\\
    \caption{The diagrams for generic split helicity amplitudes: part II.}
    \label{fig10b}
   \end{figure}

The results for MHV and $\overline{MHV}$ amplitudes indicates that
for split helicity amplitudes a better choice is to choose $\cz$'s
for the positive helicity gluons and $\cw$'s for the negative
helicity gluons. Let us make this choice and compute, as examples,
the six-gluon NMHV amplitude $A_6(+++---)$ and seven-gluon NNMHV
(while $\overline{NMHV}$) amplitude $A_7(+++----)$. The diagrammatic
representation of the computations and the results are put in
Fig.~\ref{fig8} and Fig.~\ref{fig9} respectively. Notice that in
Fig.~\ref{fig9}, we meet another situation that some triangles with
$+$ signs are combined into a domain. The helicities of the external
gluons in this domain is neither $MHV$ or $\overline{MHV}$. we call
this domain type II$+$ domain. Similarly we will meet type II$-$
domain made up of triangles with $-$ signs in the diagrams for other
amplitudes.

   \begin{figure}[ht]
    \epsfxsize=120mm%
    \hfill\epsfbox{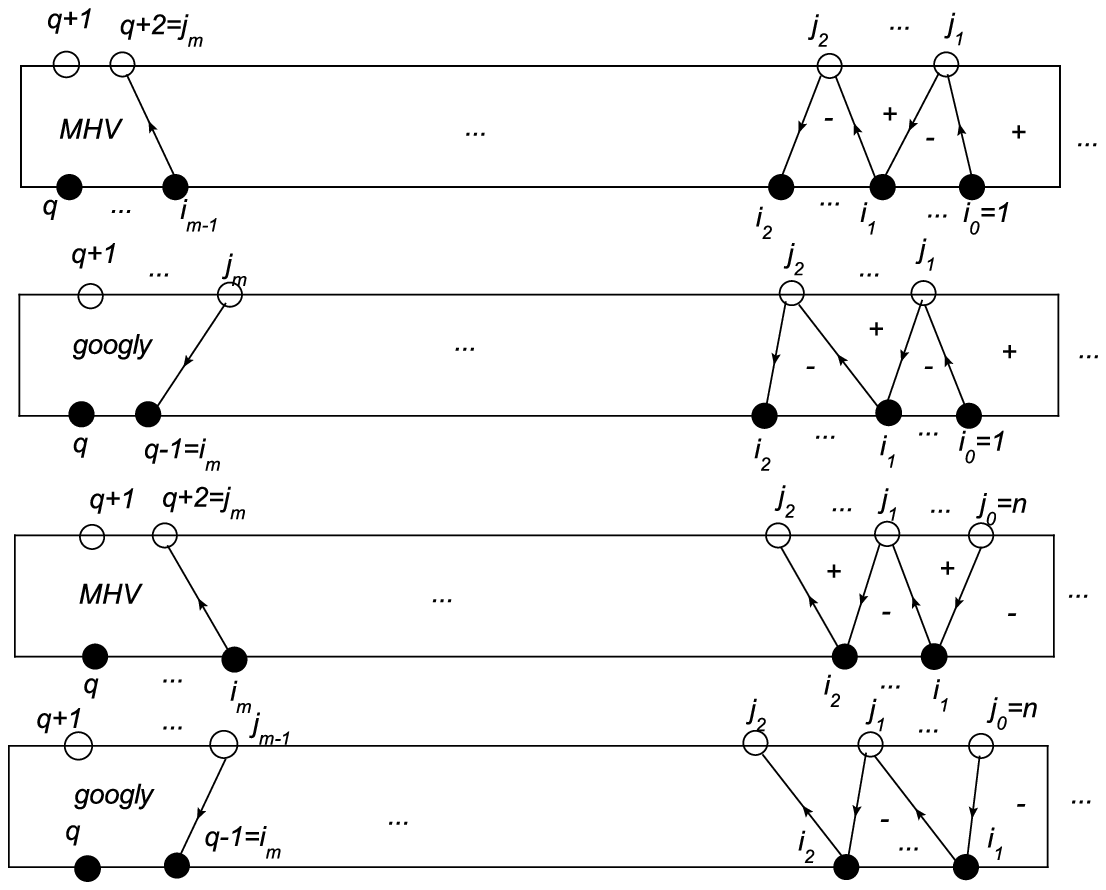}\hfill~\\
    \caption{The ziazag diagrams.}
    \label{fig11}
   \end{figure}

\subsection{General Results\label{ss34}}

Now we can draw the diagrams for the generic split helicity
amplitudes as in Fig~\ref{fig10} and Fig.~\ref{fig10b}. In these
diagrams, all domains besides the MHV or googly domains are type II
domains. We just use $+$ or $-$ to denote type II$+$ or type II$-$
domains in these diagrams. There is exactly one external gluon with
positive (negative) helicity in type II$+$ (type II$-$) domains
\footnote{The readers may wonder why we do not exchange the name for
these two domains. The reason is that  the signs for the domains
inherit from the ones of the triangles in the domains. We also treat
a single triangle as a special case of a type II domain.}. In our
diagrams, all of the nods for the gluons are choose to be white
(black) for the type II$+$ (type II$-$) domains. We can simply
express every diagram using zigzag diagrams (Fig.~\ref{fig11}).
Totally we have four types of zigzag diagrams:

\bea 1=i_0<i_1<i_2<\cdots<i_m=q-1, n>j_1>j_2>\cdots>j_m>q+1, \eea
\bea 1=i_0<i_1<i_2<\cdots<i_{m-1}<q, n>j_1>j_2>\cdots>j_m=q+2, \eea
\bea 1<i_1<i_2<\cdots<i_m<q, n=j_0>j_1>\cdots>j_m=q+2, \eea \bea
1<i_1<i_2<\cdots<i_m=q-1, n=j_0>j_1>\cdots>j_m>q+1. \eea

 The final result for the split helicity gluonic amplitudes we
want to compute is\footnote{We use $\tcm$ to emphasis that we
throw away many diagrams not contribute to the split helicity
amplitudes.}: \bea &  M^{+\cdots +-\cdots -}(Z_1, \cdots, Z_q,
W_{q+1}, \cdots, W_n)=\int \prod_{i=1}^q d^4\eta_i \prod_{j=q+1}^n
d^4\tilde\eta_j\nn\\
&  \qquad \qquad\qquad \qquad\tcm(\cz_1, \cdots, \cz_q, \cw_{q+1},
\cdots, \cw_n), \eea where \be
\tcm=\sum_{I}\tcm_I+\sum_{II}\tcm_{II}+\sum_{III}\tcm_{III}+\sum_{IV}\tcm_{IV}.
\ee The summation is over the zigzags and \bea & & \tcm_I=\int
\sd\cw_1\exp[i\cw_1\cdot\cz_1]
\tcm^+(\cw_1; \cw_{j_1}, \cdots, \cw_n)\nn\\
& & \qquad \qquad \int\sd\cz_{j_1}\exp[i\cz_{j_1}\cdot\cw_{j_1}]\int
\sd \cz_1^{\prime}\exp[i\cz_1^\pr\cdot \cw_1]\tcm^-(\cz_{j_1};
\cz_1^\pr, \cdots,
\cz_{i_1})\nn\\
& &\qquad \qquad \int \sd \cw_{j_1}^\pr \exp[i\cw_{j_1}^\pr\cdot
\cz_{j_1}]\int \sd \cw_{i_1}\exp[i\cw_{i_1}\cdot
\cz_{i_1}]\tcm^+(\cw_{i_1};  \cw_{j_2}, \cdots,
\cw_{j_1}^\pr)\nn\\
& & \qquad \qquad \cdots \int\sd \cz_{i_{m-1}}^\pr
\exp[i\cz_{i_{m-1}}\cdot \cw_{i_{m-1}}]  \int\sd
\cz_{q+2}\exp[i\cz_{q+2}\cdot \cw_{q+2}]\nn\\
& & \qquad \qquad
\int\sd\cz_{q+1}\exp[i\cz_{q+1}\cdot\cw_{q+1}]\tcm_{MHV}(\cz_{q+1},
\cz_{q+2}; \cz_{i_{m-1}}^\pr, \cdots, \cz_q),\eea

\bea & &\tcm_{II}=\int
\sd\cw_1\exp[i\cw_1\cdot\cz_1]\tcm^+(\cw_1; \cw_{j_1}, \cdots,\cw_n)\nn\\
&&\qquad \qquad
\cdots\int\sd\cw_{j_m}^\pr\exp[i\cw_{j_m}^\pr\cdot\cz_{j_m}]\int\sd\cw_{q-1}\exp[i\cw_{q-1}\cdot\cz_{q-1}]
\nn\\ & &\qquad \qquad \int\sd\cw_q
\exp[i\cw_q\cdot\cz_q]\tcm_{\overline{MHV}}(\cw_{q-1}, \cw_q;
\cw_{q+1}, \cdots, \cw_{j_{m-1}}^\pr), \eea

\bea & &\tcm_{III}=\int\sd\cz_n\exp[i\cz_n\cdot\cw_n]\tcm^-(\cz_n;
\cz_1, \cdots, \cz_{i_1})
\nn\\
& &\qquad \qquad\int\sd \cw_n^\pr \exp[i \cw_n^\pr\cdot
\cz_n]\int\sd\cw_{i_1}\exp[i\cw_{i_1}\cdot\cz_{i_1}]\tcm^+(\cw_{i_1};
\cw_{j_1}, \cdots,
\cw_n)\nn\\
& &\qquad \qquad
\cdots\int\sd\cz_{i_m}^\pr\exp[i\cz_{i_m}^\pr\cdot\cw_{i_m}]
\int\sd\cz_{q+2}\exp[i\cz_{q+2}\cdot\cw_{q+2}]\nn\\
&&\qquad \qquad
\int\sd\cz_{q+1}\exp[i\cz_{q+1}\cdot\cw_{q+1}]\tcm_{MHV}(\cz_{q+1},
\cz_{q+2}; \cz_{i_m}^\pr, \cdots, \cz_q),
 \eea

\bea &
&\tcm_{IV}=\int\sd\cz_n\exp[i\cz_n\cdot\cw_n]\tcm^-(\cz_n; \cz_1, \cdots, \cz_{i_1})\nn\\
&&\qquad \qquad\cdots\int\sd\cw_{j_{m-1}}^\pr\exp[i\cw_{j_{m-1}}^\pr\cdot\cz_{j_{m-1}}]\int\sd\cz_{q-1}\exp[i\cz_{q-1}\cdot\cw_{q-1}]\nn\\
& &\qquad \qquad
\int\sd\cw_q\exp[i\cw_q\cdot\cz_q]\tcm_{\overline{MHV}} (\cw_{q-1},
\cw_q; \cw_{q+1} \cdots, \cw_{j_{m-1}}^\pr).\eea

Here the expressions for the type II domain are:

\bea \tcm^+(\cw_j; \cw_i, \cw_{i+1}, \cdots,
\cw_l)&=&(-1)^{l-i}\prod_{k=i}^{l-1}
\tcm^+(\cw_j, \cw_k, \cw_{k+1}),\\
\tcm^-(\cz_j; \cz_i, \cz_{i+1}, \cdots,
\cz_l)&=&(-1)^{l-i}\prod_{k=i}^{l-1}\tcm^-(\cz_j, \cz_k, \cz_{k+1}).
\eea

The expressions for MHV and $\overline{MHV}$ domains are: \bea
\tcm_{MHV}(\cz_{q+1},\cz_{q+2}; \cz_{i_m}, \cdots,
\cz_q)&=&(-1)^{q-i_m+1}\prod_{k=i_m}^{q-1}M^-(\cz_{q+1}, \cz_k, \cz_{k+1})\nn\\
&\times&M^-(\cz_{q+1}, \cz_{q+2},
\cz_{i_m}), \\
\tcm_{\overline{MHV}}(\cw_{q-1}, \cw_q; \cw_{q+1}, \cdots,
\cw_{j_m})&=&(-1)^{j_m-q}\prod_{i=q+1}^{j_m-1}M^+(\cw_{q-1}, \cw_q,
\cw_{j_m})\nn\\
&\times&M^+(\cw_{q-1}, \cw_q, \cw_{j_m}),\eea where $M^-(\cz_1,
\cz_2, \cz_3), M^+(\cw_1, \cw_2, \cw_3)$ are given in
eqs.~(\ref{3ptp}, \ref{3ptn}). The fact that all the amplitudes are
composed of the domains shows that all tree-level split helicity
amplitudes are triangulable.

We can see from these result that the most natural choices of $\cw$'s and $\cz$'s for
$1$, $q$, $q+1$, $n$ are different in different terms. Similar phenomenon appears in the
amplitudes for gravitons in \cite{ACCK}.

By using the 'triangle identity' in \cite{ACCK}, we can rewrite
eqs.~(\ref{3ptp}, \ref{3ptn}) in a suitable way and perform all of
the Grassman integrations to get: \be M(Z_1, +; \cdots; Z_q, +;
W_{q+1}, -; \cdots; W_n,
-)=\sum_{I}M_I+\sum_{II}M_{II}+\sum_{III}M_{III}+\sum_{IV}M_{IV}.
\ee Here the summation is still over the zigzags. Now \bea & &
M_I=\int d^4 W_1\exp[iW_1\cdot Z_1]
\tcm^+(W_1, +; W_{j_1}, -; \cdots; W_n, -)\nn\\
& & \qquad \qquad \int d^4 Z_{j_1}\exp[iZ_{j_1}\cdot W_{j_1}]\int
d^4 Z_1^{\prime}\exp[iZ_1^\pr\cdot W_1]\tcm^-(Z_{j_1}, -; Z_1^\pr,
+; \cdots;
Z_{i_1}, +)\nn\\
& &\qquad \qquad \int d^4 W_{j_1}^\pr \exp[iW_{j_1}^\pr\cdot
Z_{j_1}]\int d^4 W_{i_1}\exp[iW_{i_1}\cdot Z_{i_1}]\tcm^+(W_{i_1},
+;  W_{j_2}, -; \cdots;
W_{j_1}^\pr, -)\nn\\
& & \qquad \qquad \cdots \int d^4 Z_{i_{m-1}}^\pr \exp[i
Z_{i_{m-1}}\cdot W_{i_{m-1}}]  \int d^4
Z_{q+2}\exp[i Z_{q+2}\cdot W_{q+2}]\nn\\
& & \qquad \qquad \int d^4 Z_{q+1}\exp[iZ_{q+1}\cdot
W_{q+1}]\tcm_{MHV}(Z_{i_{m-1}}^\pr, +; \cdots; Z_q, +; Z_{q+1}, -;
Z_{q+2}, -),\eea
\bea & &M_{II}=\int d^4W_1\exp[iW_1\cdot
Z_1]\tcm^+(W_1, +; W_{j_1}, -; \cdots; W_n,
-)\nn\\
&&\qquad \qquad \cdots\int d^4 W_{j_m}^\pr\exp[iW_{j_m}^\pr\cdot
Z_{j_m}]\int d^4W_{q-1}\exp[iW_{q-1}\cdot Z_{q-1}] \nn\\ & &\qquad
\qquad \int d^4 W_q \exp[iW_q\cdot
Z_q]\tcm_{\overline{MHV}}(W_{q-1}, +; W_q, +; W_{q+1}, -; \cdots;
W_{j_{m-1}}^\pr, -), \eea
\bea & &M_{III}=\int d^4Z_n\exp[iZ_n\cdot
W_n]\tcm^-(Z_n, -; Z_1, +; \cdots, Z_{i_1}, +)
\nn\\
& &\qquad \qquad\int d^4 W_n^\pr \exp[i W_n^\pr\cdot Z_n]\int
d^4W_{i_1}\exp[iW_{i_1}\cdot Z_{i_1}]\tcm^+(W_{i_1}, +; W_{j_1}, -;
\cdots,
W_n, -)\nn\\
& &\qquad \qquad \cdots\int d^4Z_{i_m}^\pr\exp[iZ_{i_m}^\pr\cdot
W_{i_m}]
\int d^4Z_{q+2}\exp[iZ_{q+2}\cdot W_{q+2}]\nn\\
&&\qquad \qquad \int d^4Z_{q+1}\exp[iZ_{q+1}\cdot
W_{q+1}]\tcm_{MHV}(Z_{i_m}^\pr, +; \cdots; Z_q, +; Z_{q+1}, -;
Z_{q+2}, -),
 \eea
\bea & &\tcm_{IV}=\int d^4Z_n\exp[iZ_n\cdot W_n]\tcm^-(Z_n, -; Z_1,
+;
\cdots, Z_{i_1}, +)\nn\\
&&\qquad \qquad\cdots\int d^4W_{j_{m-1}}^\pr\exp[iW_{j_{m-1}}^\pr\cdot Z_{j_{m-1}}]\int d^4Z_{q-1}\exp[iZ_{q-1}\cdot W_{q-1}]\nn\\
& &\qquad \qquad \int d^4W_q\exp[iW_q\cdot Z_q]\tcm_{\overline{MHV}}
(W_{q-1}, +; W_q, +; W_{q+1}, -; \cdots; W_{j_{m-1}}^\pr, -).\eea

The express for the type II domains and MHV/$\overline{MHV}$ domains
are: \be \tcm^-(Z_j, -; Z_i, +; Z_{i+1}, +; \cdots; Z_l,
+)=(-1)^{l-i}\prod_{k=i}^{l-1}\tcm^-(Z_j, Z_k, Z_{k+1}) \ee \be
\tcm^+(W_j, +; W_i, -; W_{i+1}, -; \cdots; W_l,
-)=(-1)^{l-i}\prod_{k=i}^{l-1}\tcm^+(W_j, W_k, W_{k+1}), \ee \bea
&&\tcm_{MHV}(Z_{i_m}, +; \cdots; Z_q, +; Z_{q+1}, -; Z_{q+2},
-)=(-1)^{q-i_m}\tcm^-(Z_{q-1}, Z_{q+1}, Z_{q+2})\nn\\&& \tcm^-(Z_q,
Z_{q-1}, Z_{q+1})\prod_{l=i_m}^{q-2}\tcm^-(Z_l, Z_{l+1},
Z_{q+2}),\eea \bea &&\tcm_{\overline{MHV}}(W_{q-1}, +; W_q, +;
W_{q+1}, -;\cdots; W_{j_m}, -)=(-1)^{j_m-q+1}\tcm^+(W_q, W_{q+1},
W_{q+2})\nn\\&&\tcm^+(W_{q-1}, W_q,
W_{q+2})\prod_{l=q+2}^{j_m-1}M(W_{q-1}, W_l, W_{l+1})\eea

The three gluon amplitudes can be computed using eq.~(\ref{sgn})
\cite{Mason:2009sa}: \bea & & \tcm^-(Z_j, Z_k, Z_{k+1})\nn\\&=&\int
d^4W_k \exp(iW_k\cdot Z_k)\sgn(W_k\cdot Z_j)\sgn(W_k\cdot
Z_{k+1})\sgn(Z_{k+1}IZ_j)\nn\\
&=&\frac{\d^2(\lan k+1, j\ran \mu_k+\lan k, k+1\ran \mu_j+\lan j,
k\ran \mu_{k+1})}{\lan j, k \ran \lan k, k+1\ran \lan k+1, j\ran}.
\eea
\bea & &\tcm^+(W_j, W_k, W_{k+1})\nn\\
&=&\int d^4Z_{k+1}\exp(iZ_{k+1}\cdot W_{k+1})\sgn(Z_{k+1}\cdot
W_k)\sgn(Z_{k+1}\cdot W_j)\sgn(W_jIW_k)\nn\\
&=&\frac{\d^2([k+1, j]\tilde\mu_k+[k, k+1]\tilde\mu_j+[j,
k]\tilde\mu_{k+1})}{[j, k][k, k+1][k+1, j]} \eea

\section{The computations of six-gluon NMHV amplitudes\label{sec4}}

   \begin{figure}[ht]
    \epsfxsize=100mm%
    \hfill\epsfbox{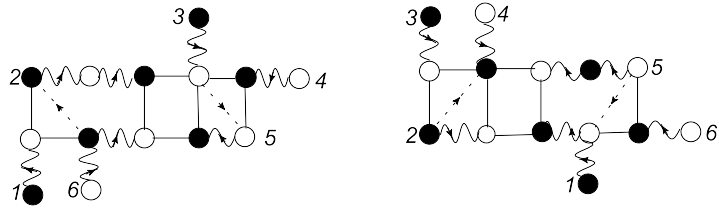}\hfill~\\
    \caption{The Hodges diagrams obtained from Fig.~9.}
    \label{fig63}
   \end{figure}

   \begin{figure}[ht]
    \epsfxsize=100mm%
    \hfill\epsfbox{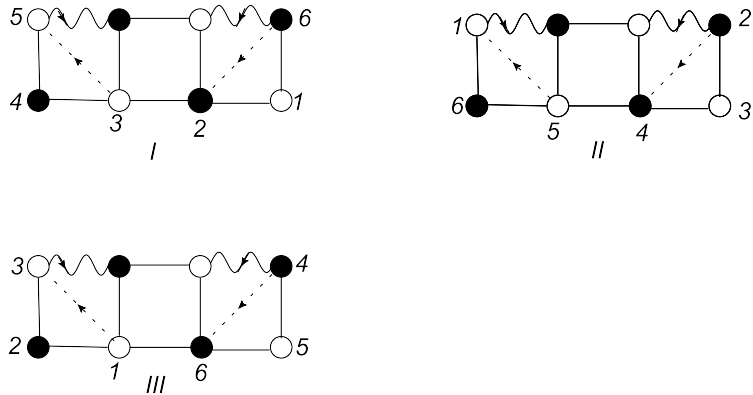}\hfill~\\
    \caption{The Hodges diagrams for $6$-gluon NMHV amplitude obtained from \cite{ACCK}.}
    \label{fig61}
   \end{figure}

   \begin{figure}[ht]
    \epsfxsize=100mm%
    \hfill\epsfbox{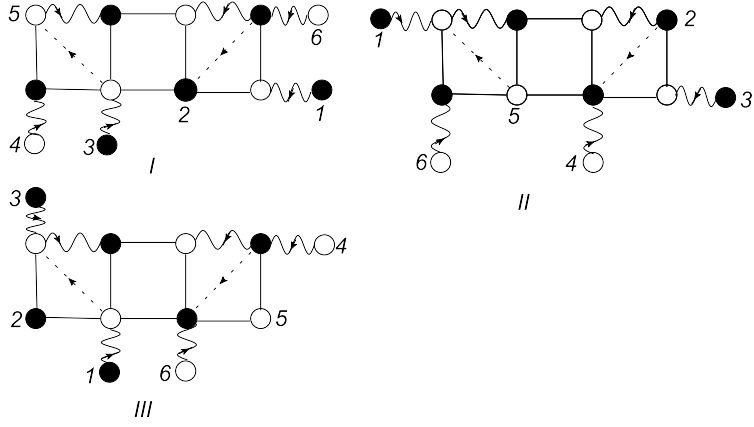}\hfill~\\
    \caption{The Hodges diagrams from the twistor transformation of Fig.~15.}
   \label{fig62}
   \end{figure}

As an example, now we compute the six-gluon split next-to-MHV (NMHV)
amplitude $M(1^+, 2^+, 3^+, 4^-, 5^-, 6^-)$ and show that we
reproduce the old result in momentum space. The Hodges diagrams has
been given in Fig.~\ref{fig8}. We can expand these diagrams as in
Fig.~\ref{fig63}. In \cite{ACCK}, Hodges diagrams for $6$-gluon NMHV
amplitudes are given. After performing parity transformation, the
diagrams are listed in Fig.~\ref{fig61}. To compare with our
diagrams, we perform a further twistor space transformation as in
Fig.~\ref{fig62}. It is not hard to seen that the two diagrams in
Fig.~\ref{fig63} are the same as the first two diagrams in
Fig.~\ref{fig62}. We will show in the following that diagram (III)
in Fig.~\ref{fig62} will not contribute to this split NMHV amplitude
as expected from our general discussions (For this, it is enough to
show that the contribution from the third diagram in
Fig.~\ref{fig61} vanishes). We will also show that the first two
diagrams in Fig.~\ref{fig61} will reproduce the momentum space
results. This will in turn confirm that our results give the correct
momentum space results.

First we have the following Grassman integration: \bea &&
M^{3+3-}(W_1, Z_2, W_3, Z_4, W_5, Z_6)=\int
d^4\eta_2d^4\tilde\eta_5\nn\\
&& \qquad \qquad M(\cw_1, \cz_2, \cw_3, \cz_4, \cw_5,
\cz_6)|_{\tilde\eta_1=\tilde\eta_3=\eta_4=\eta_6=0}. \eea From now
on, unless otherwise claimed, we use the convention that small $i$
takes odd integer values $1, 3, 5$ and capital $J$ takes even
integer value $2, 4, 6$. The contribution from diagram (III) in
Fig.~\ref{fig61} is \bea \int d^4\eta_2d^4\tilde\eta_5 [M^+(\cw_1,
\cz_2, \cw_3) M_4(\cw_3, \cz_4, \cz_6,
 \cw_1) M^-(\cz_4, \cw_5, \cz_6)]|_{\tilde\eta_1=\tilde\eta_3=\eta_4=\eta_6=0}. \eea
Now let us consider the integration over $\eta_2$, noticing that
$\cz_2$ only appears in $M^+(\cw_1, \cz_2, \cw_3)$  through the
factors $\sgn(\cw_1\cdot\cz_2)$ and $\sgn(\cw_3\cdot\cz_2)$. But
after imposing $\tilde\eta_1=\tilde\eta_3=0$, $\eta_2$'s do not
appear in these factors. So this integral vanishes and this diagram
does not contribute to the six-gluon split NMHV amplitudes as
claimed.

Now we turn to consider the contribution from diagram (I) in
Fig.~\ref{fig61} to $M^{3+3-}(W_i, Z_J)$, which is \bea  \int
d^4\eta_2d^4\tilde\eta_5 [M^+(\cw_3, \cz_4, \cw_5) M_4(\cw_3,
\cw_5, \cz_6,
 \cz_2)M^-(\cw_1, \cz_2, \cz_6)]|_{\tilde\eta_1=\tilde\eta_3=\eta_4=\eta_6=0}.\label{6pt1} \eea
We can see that after imposing
$\tilde\eta_1=\tilde\eta_3=\eta_4=\eta_6=0$, there will be no
$\eta_2$'s and $\tilde\eta_5$'s in $M^+(\cw_3, \cz_4, \cw_5)$ and
$M^-(\cw_1, \cz_2, \cz_6)$. So the above equation equals to \bea
M^+(W_3, Z_4, W_5)M^-(W_1, Z_2, Z_6)\int d^4\eta_2d^4\tilde\eta_5
[ M_4(\cw_3, \cw_5, \cz_6,
 \cz_2)]|_{\tilde\eta_1=\tilde\eta_3=\eta_4=\eta_6=0}.\eea
Now we use the link representation for $M_4(\cw, \cw, \cz, \cz)$
in \cite{ACCK} to get: \be M_4(\cw_3, \cw_5, \cz_6,
 \cz_2)]=\int dc_{iJ}\frac{\exp[ic_{iJ}\cw_i\cdot\cz_J]}{c_{36}c_{52}(c_{36}c_{52}-c_{32}c_{56})},\ee where $i=3, 5$ and
$J=6, 2$. Then we have \bea \int d^4\eta_2d^4\tilde\eta_5 [
M_4(\cw_3, \cw_5, \cz_6,
 \cz_2)]|_{\tilde\eta_1=\tilde\eta_3=\eta_4=\eta_6=0}=
 \int dc_{iJ}\frac{c_{52}^3\exp[ic_{iJ}W_i\cdot Z_J]}{c_{36}(c_{36}c_{52}-c_{32}c_{56})}
 \eea
We also use the following link representation of three-gluon
amplitudes: \be M_{-}(W_1, Z_2, Z_6)=\sgn(\lan 2, 6\ran)\int
dc_{12}dc_{16}\frac{\exp[i(c_{12}W_1\cdot Z_2+c_{16}W_1\cdot
Z_6)]}{c_{12}c_{16}}, \ee \be M_{+}(W_3, Z_4, W_5)=\sgn([5,
3])\int dc_{34}dc_{54}\frac{\exp[i(c_{34}W_3\cdot
Z_4+c_{16}W_3\cdot Z_4)]}{c_{34}c_{54}}. \ee

We conclude that the contribution from diagram (I) is \bea &
&\sgn(\lan 2, 6\ran [5, 3])\int\prod_{(i, J)=(odd, \, even)\ne(1,
4)}dc_{iJ}\exp[i c_{iJ}W_i\cdot Z_J]\nn\\&&\qquad \qquad
\qquad\qquad
\frac{c_{52}^3}{c_{54}c_{34}c_{16}c_{12}c_{36}(c_{52}c_{36}-c_{56}c_{32})}.\label{link}
\eea

Now to compute this amplitude in momentum space, we need to
perform the following transformation: \bea M^{3+3-}(\lam_i,
\tilde\lam_i)&=&\int\prod_{{i\,
odd}}d^2\tilde\mu_i\exp[-i\tilde\mu_i\lam_i]\int\prod_{J\,even}d^2\mu_J\exp[-i\mu_J\tilde\lam_J]\nn\\
&&M^{3+3-}(W_1, Z_2, W_3, Z_4, W_5, Z_6) \eea Similar to the
computations in \cite{ACCK}, we get that the contribution from
diagram (I) is: \bea M^{3+3-}_I(\lambda_i,
\tilde\lam_i)=\frac{\lan6|(p_1+p_2)|3]^3\d^4(\sum_i\lam_i\tilde\lam_i)
}{\lan 6, 1\ran \lan 2, 1\ran [3, 4] [5, 4]\lan 2|(p_6+p_1)|5]
(p_6+p_1+p_2)^2}. \eea

As to the contribution from diagram (II) in Fig.~\ref{fig61},
after getting the link representation as in eq.~(\ref{link}), we
find that this can be obtained from eq.~(\ref{link}) by exchange
$1$ and $3$, $4$ and $6$. Then this directly gives the results.

Finally we get the following results in the momentum space: \bea
&& M(1^+, 2^+, 3^+, 4^-, 5^-, 6^-)\nn\\
&&\qquad \qquad=
\frac{\d^4(\sum_i\lam_i\tilde\lam_i)}{\lan2|p_3+p_4|5]}  \left(
\frac{\lan 4|p_2+p_3|1]}{\lan 2, 3\ran \lan 3, 4\ran [5, 6] [6,
1](p_2+p_3+p_4)^2}\right.\nn\\
&&\qquad \qquad \qquad +\left.\frac{\lan 6 |p_4+p_5|3]}{\lan 6,
1\ran \lan 1, 2\ran [3, 4] [4, 5] (p_3+p_4+p_5)^2} \right).\eea
This reproduce the correct momentum space results.

\section{Conclusion and Discussions}

In this paper, we studied all tree-level split helicity amplitudes
in ambi-twistor space in details. Using Grassman integration, we
threw away many diagrams which do not contribute to these special
helicity configurations. It is interesting to see whether the
similar simplification could happen in other cases. We also found
a way to organize the remaining diagrams. We found that all of
these remaining diagrams could be divided into triangles in a
suitable way. Similar structure was found in \cite{talkKaplan,
talkNima} for the tree-level amplitudes with up to $15$ external
particles. Even though we only considered the amplitudes with
fixed helicity, our result gave strong support to the claim that
all the tree-level amplitudes could be
triangulable\cite{talkKaplan,talkNima}.

In our study, we found that the diagrams with non-vanishing
contribution could be classified into four kinds of zigzag pattern.
These zigzag diagrams are reminiscent of the similar diagram
appeared in the momentum space results for the split helicity
amplitudes studied in \cite{BFRSV}. 
We are curious to understand the relation between these two kinds
of zigzag diagrams. It is also interesting to study how the dual
conformal covariance \cite{Drummond:2008vq} of these split
helicity amplitudes is manifest in this framework.

In our computations of the six-gluon NMHV amplitude, the link
representation \cite{ACCK} of the amplitudes is quite helpful. We
expect that the link representation will play a similar role in the
computations of more general amplitudes. We also hope that our study
here will be useful for the studies of all of the tree-level
amplitudes in ${\cal N}=4$ super Yang-Mills, and in ${\cal N}=8$
supergravity in ambitwistor space. These amplitudes in momentum
space were computed in \cite{DrummondHenn} and \cite{DSVW}. Some
studies on the amplitudes in twistor space have been performed in
\cite{Mason:2009sa}.

We found that, opposite to our intuition based on the weight of
the amplitudes under the scaling of $\cw$ and $\cz$, it is more
convenient in the analysis to choose $\cz$ for almost all of the
positive helicity gluons and $\cw$ for almost all of the negative
helicity gluons for the general split helicity amplitudes. However
we also saw that in the practical computations of the six-gluon
split NMHV amplitude, it is more convenient to choose $\cw$'s and
$\cz$'s alternately as in \cite{ACCK}. 
It is not clear if this is always the case. The advantage of our
choice is that we could easily determine the diagrams with
non-vanishing contribution.

\section*{Acknowledgments}
The work was partially supported by NSFC Grant No.10535060,
10775002 and NKBRPC (No. 2006CB805905). BC would like to thank
KITPC for hospitality, where part of the work was done. JW would
like to thank Matteo Bertolini and International School for
Advanced Studies (SISSA) for hospitality. JW would also like to
thank Peng Gao, Johannes Henn, and Radu Roiban for very helpful
discussions. The Feynman diagrams are drawn using Jaxodraw
\cite{Jaxodraw1, Jaxodraw2}, which is based on Axodraw
\cite{axodraw}.

\end{document}